\documentclass[twocolumn,aps,prd,amsmath,amssymb,floatfix,amsmath,superscriptaddress]{revtex4-2}

\usepackage{graphicx}% Include figure files
\usepackage{dcolumn}% Align table columns on decimal point
\usepackage{bm}% bold math
\usepackage{braket}
\usepackage{color,epsfig}
\usepackage{bm}
\usepackage{slashed}
\usepackage{hyperref}
\usepackage{subfigure}
\usepackage{feynmf}
\usepackage{calrsfs}

\usepackage{booktabs}
\usepackage{siunitx}

\newcommand{\bea}{\begin{eqnarray}}
\newcommand{\eea}{\end{eqnarray}}
\newcommand{\be}{\begin{equation}}
\newcommand{\ee}{\end{equation}}

\renewcommand\vec{\bm}

\begin{document}

\preprint{APS/123-QED}
\title{Very Special Linear Gravity: A Gauge-Invariant Graviton Mass}
%\title{VSR Linearized Gravity:\\A Gauge-Invariant Graviton Mass}% Force line breaks with \\

\author{Jorge Alfaro}
 \altaffiliation{jalfaro@fis.puc.cl}%Lines break automatically or can be forced with \\
\author{Alessandro Santoni}%
 \email{asantoni@uc.cl}
\affiliation{Instituto de Física, Pontificia Universidad de Católica de Chile, \\
 Avenida Vicuña Mackenna 4860, Santiago, Chile 
}%

\begin{abstract}
Linearized gravity in the Very Special Relativity (VSR) framework is considered. We prove that this theory allows for a non-zero graviton mass $m_g$ without breaking gauge invariance nor modifying the relativistic dispersion relation. We find the analytic solution for the new equations of motion in our gauge choice, verifying as expected the existence of only two physical degrees of freedom. Finally, through the geodesic deviation equation, we confront some results for classic gravitational waves (GW) with the VSR ones: we see that the ratios between VSR effects and classical ones are proportional to $(m_g/E)^2$, $E$ being the energy of a graviton in the GW. For GW detectable by the interferometers LIGO and VIRGO this ratio is at most $10^{-20}$. However, for GW in the lower frequency range of future detectors, like LISA, the ratio increases significantly to $ 10^{-10}$, that combined with the anisotropic nature of VSR phenomena may lead to observable effects.

\end{abstract}

\maketitle

%\tableofcontents

\section{Introduction}

In 2006, A. Cohen and S. Glashow presented for the first time the idea of Very Special Relativity (VSR), where they imagined to restrict space-time invariance to a subgroup of the full Lorentz group \cite{cohen2006very}, usually the subgroup $SIM(2)$. The advantage of this theory is that, while it does not affect the classical prediction of Special Relativity, it can explain the existence of neutrino masses \cite{cohen2006lorentz} without the addition of new exotic particles or tiny twisted space dimensions, which until now have not been observed in experiments. 

The addition of either $P$, $CP$, or $T$ invariance to $SIM(2)$ symmetry enlarges the entire symmetry group again to the whole Lorentz group. That implies the absence of VSR effects in theories where one of the above three discrete transformations is conserved. 

Now, since we know thanks to Sakharov conditions \cite{Sakharov:1967dj} that these discrete symmetries must be broken in cosmology, the effects of VSR in this framework become worthy of being studied. 

Therefore, in this paper, we aim to construct a $SIM(2)$-invariant version of linearized gravity, describing the dynamics of the space-time perturbation field $h_{\mu \nu}$. Such a theory may be used as a starting point for the study of VSR consequences in the propagation of gravitational waves in a Lorentz breaking background. 

In the end, our analysis will correspond to a massive graviton model. That could be of great interest due to the various recent applications that are being explored for massive gravity, from dark matter \cite{aoki2016massive} to cosmology \cite{comelli2012perturbations}, despite the strong boundaries we already have on the graviton mass \cite{de2017graviton}. 

Until now, massive gravity models were usually constructed as Lorentz invariant \cite{mgrav1,mgrav2}. Nevertheless, as in the case of Electromagnetism and the Proca Theory, there is no way of trivially preserving both Lorentz and Gauge invariance when giving mass to the graviton. 

Giving up on the Gauge invariance directly leads to the appearance of three additional degrees of freedom (D.o.F.) respect to the ones of General Relativity (GR), which are responsible for different pathologies of these theories, like the vDVZ discontinuity \cite{disc1,disc2} and ghost modes (i.e. the Boulware-Deser ghost \cite{ghost}). Many of these problems have already been solved with the Vainshtein Mechanism \cite{vain} and the fine-tuned dRGT action \cite{dRGT} to avoid ghosts, making dRGT massive gravity a good candidate to solve the cosmological constant problem. Even so, dealing with cosmology brings up new problems and instabilities which have not already been solved \cite{cosmo}.

Giving up on Lorentz invariance, that is what we are considering in this paper by implementing VSR, is the other viable possibility for massive gravity. Experience with VSR Electrodynamics \cite{vsrqed} and VSR massive Neutrinos \cite{cohen2006lorentz} tell us that VSR extensions avoid the introduction of ghosts in the spectrum. In fact, as we will see, gauge invariance of our formulation does not allow for new additional D.o.F. other than the usual two of the massless graviton, getting around most of the problems cited above. Nevertheless, these advantages come at the price of considering new non-local terms in the theory and assuming a preferred space-time null direction, represented by the lightlike four-vector $n^\mu$.

%The paper is organized as follows: in Sec. \ref{sec1}, we start by finding the most general lagrangian for the field $h_{\mu \nu}$ in the VSR framework, from which derive the equations of motion (E.o.M). In Sec. \ref{sec2}, we use all of our gauge freedom to fix a consistent set of conditions for the field $h_{\mu\nu}$, so that the E.o.M assume a massive Klein-Gordon form. In Sec. \ref{sec3}, we find an explicit solution for the components of $h_{\mu \nu}$ in function of the just two physical degree of freedom. In Sec. \ref{sec4}, we verify the correctness of the massless limit of our theory. Finally, in Sec. \ref{sec5}, we apply the geodesic deviation equation for the gravitational waves case but in the framework of VSR, confronting it with the “classical" results and formulating hypothesis for an upper bound to the VSR effects.\\

\section{Lagrangian of the Field $h_{\mu \nu}$}
First of all, we want to find the general expression for the lagrangian quadratic in the graviton field $h_{\mu \nu} $
\begin{eqnarray} \label{lagr} \centering
   && \mathcal{L}_g = \frac12\, h^{\mu \nu} O_{\mu \nu \alpha \beta} h^{\alpha \beta} \, .
\end{eqnarray}
To construct the most generic form of the operator $O$, we restricted our search to terms containing up to two
derivatives and parameters with dimension up to four in energy. In momentum space, the objects we can put together to build up terms for $O_{\mu \nu \alpha \beta}$ are: $\eta_{\mu \nu}$, $ p_\mu$ and $N_\mu = \frac{n_\mu}{n \cdot p}$. Therefore, symbolically speaking, the operator $O$ assumes the structure
\begin{equation} \label{genO}
    O =  3 \, \eta \eta + 9 \,  p p \eta +12 \, p N \eta + 12 \, p p N N \ , 
\end{equation}
where the numbers indicate how many different terms we can construct with each set of objects. 

\subsection{Parameters' Restrictions}
 To restrict the parameters' phase space of the model, we first impose gauge invariance under the same transformations group of $h_{\mu\nu }$ obtained in linearized GR, which in momentum space reads as
\begin{equation}
    \delta h_{\mu\nu}= p_\mu \xi_\nu +p_\nu \xi_\mu \,.
\end{equation}
For our Lagrangian \eqref{lagr}, this gauge invariance translates in the condition $O_{\mu \nu \alpha \beta} p^\alpha = 0$. Furthermore, we must impose for $O$ the index symmetries $\mu \Longleftrightarrow \nu$, $\alpha \Longleftrightarrow \beta$, $\mu \nu \Longleftrightarrow \alpha \beta$.

This way, we reduce the 36 parameters of the expression \eqref{genO} to only two free parameters: a constant global factor $\chi$, that we can identify with the Einstein-Hilbert constant $\chi = \frac{c^4}{16 \pi G}$, and $m^2_g$, that has dimensions of a mass squared and, as we will see, plays the role of a graviton mass.
\begin{widetext}

\begin{equation}
    \begin{aligned}
    O_{\mu \nu \alpha \beta} = & \;\;
     % pp-sector
    \chi \; \bigg ( p_\mu p_\nu \eta_{\alpha \beta}-\frac{1}{2} p_\mu p_\alpha \eta_{\nu \beta} - \frac{1}{2} p_\mu p_\beta \eta_{\nu \alpha } + p_\alpha p_\beta \eta_{\mu \nu } -\frac{1}{2} p_\nu p_\beta \eta_{\mu \alpha } -\frac{1}{2} p_\nu p_\alpha \eta_{\mu \beta } \\
    & - p^2 \eta_{\mu \nu} \eta_{\alpha \beta }+ \frac{1}{2} p^2 \eta_{\mu \alpha} \eta_{\nu \beta} + \frac{1}{2} p^2 \eta_{\mu \beta} \eta_{\nu \alpha}
    % g-sector
     +m^2_g \eta_{\mu \nu} \eta_{\alpha \beta }- \frac{m^2_g}{2} \eta_{\mu \alpha} \eta_{\nu \beta} -\frac{m^2_g}{2} \eta_{\mu \beta} \eta_{\nu \alpha}  \\
    % ppNN-sector
    & - m^2_g N_\mu N_\nu p_\alpha p_\beta +\frac{m^2_g}{2} N_\mu N_\alpha p_\nu p_\beta + \frac{m^2_g}{2} N_\mu N_\beta p_\nu p_\alpha  +\frac{m^2_g}{2} N_\nu N_\alpha p_\mu p_\beta +\frac{m^2_g}{2} N_\nu N_\beta p_\mu p_\alpha - m^2_g N_\alpha N_\beta p_\mu p_\nu \\ 
    & + m^2_g p^2 N_\mu N_\nu g_{\alpha \beta } - \frac{m^2_g}{2} p^2 N_\mu N_\alpha g_{\nu \beta} - \frac{m^2_g}{2} p^2 N_\mu N_\beta g_{\nu \alpha}
    - \frac{m^2_g}{2} p^2 N_\nu N_\beta \eta_{\mu \alpha} - \frac{m^2_g}{2} p^2 \eta_{\mu \beta} N_\nu N_\alpha + m^2_g p^2 N_\alpha N_\beta \eta_{\mu \nu}\\
    % pN-sector
    & - m^2_g \eta_{\mu \nu} N_\alpha p_\beta - m^2_g \eta_{\mu \nu} p_\alpha N_\beta +\frac{m^2_g}{2} \eta_{\mu \alpha} N_\nu p_\beta +\frac{m^2_g}{2} \eta_{\mu \alpha} p_\nu N_\beta +\frac{m^2_g}{2} \eta_{\mu \beta} N_\nu p_\alpha +\frac{m^2_g}{2} \eta_{\mu \beta} p_\nu N_\alpha \\
    & +\frac{m^2_g}{2} \eta_{\nu \alpha} N_\mu p_\beta +\frac{m^2_g}{2} \eta_{\nu \alpha} p_\mu N_\beta +\frac{m^2_g}{2} \eta_{\nu \beta} N_\mu p_\alpha +\frac{m^2_g}{2} \eta_{\nu \beta} p_\mu N_\alpha - m^2_g \eta_{\alpha \beta } N_\mu p_\nu - m^2_g \eta_{\alpha \beta } p_\mu N_\nu  \bigg ) \, .
    \end{aligned}
\end{equation}

\end{widetext}

\subsection{Equations of Motion (E.o.M.)}
At this point, varying the expression of the Lagrangian \eqref{lagr} respect to $h_{\mu \nu}$, we obtain the E.o.M for the graviton field in momentum space, which reads as: 
\begin{equation}\label{eomh}
   O_{\mu \nu \alpha \beta} h^{\alpha \beta}  = 0 \,.
\end{equation}
In the following, we will use the definition $h = \eta^{\mu \nu} h_{\mu \nu}$, to simplify the notation. \\

\section{Gauge Choices}

Now, we want to fix a gauge for $ h_{\mu \nu}$ so that we get to an E.o.M like Klein-Gordon, and therefore a classic massive dispersion relation for the graviton.

Firstly, we use the linearized diffeomorphism gauge freedom to fix a Lorentz gauge $p^\mu  h_{\mu \nu} = 0$. This way, using the Lorentz condition and contracting the E.o.M with $p_\mu$, $n_\mu$ and $\eta_{\mu \nu}$ we get to the two useful equations
\begin{eqnarray}
   && p^2 h= 0\, , \label{eqh}\\
   && p_\nu h + p^2 N^\mu h_{\mu \nu} = 0\, \label{eqph} .
\end{eqnarray}

\subsection{Additional Gauge Freedom}

We observe that the Lorentz gauge does not fix the gauge uniquely, in fact, we can still make a new gauge transformation satisfying
\begin{equation}
    p^\mu h'_{\mu \nu} = 0 = p^\mu h_{\mu \nu} + p^2 \xi_\nu + p\cdot \xi p_{\nu} = p^2 \xi_\nu + p\cdot \xi p_{\nu} \, ,
\end{equation}
from which we derive the form of $\xi_{\nu}$
\begin{equation} \label{eqxi}
    \xi_\nu = ( c_\nu- \frac{1}{2}\frac{p\cdot c}{p^2}p_\nu ) \delta(p^2) \, ,
\end{equation}
with $c_\nu$ arbitrary, meaning we still have the freedom to impose at least four conditions on $h_{\mu \nu}$ by fixing $c_\mu$. \\

\subsection{Fixing $h=0$}
At this point, we want to use this additional gauge freedom to fix $h= 0$. From \eqref{eqh} we find for $h$ the solution $h = h_0 \delta(p^2)$. Therefore, we make a new gauge transformation satisfying \eqref{eqxi} to get in a gauge where
\begin{equation}
    h'=0 = h + 2 p^\mu \xi_\mu = \delta(p^2) (h_0 + p \cdot c) \, ,
\end{equation}
which always has at least the solution
\begin{equation} \label{eqcp}
    c \cdot p = -h_0 \, .
\end{equation}
Therefore, with such a choice, we can always impose the traceless condition $h=0$. \\

\subsection{Fixing $n^\mu h_{\mu \nu}= 0$}
With the additional condition $h=0$, equation \eqref{eqph} reduces to
\begin{equation}
    p^2 N^\mu h_{\mu \nu} = 0 \, ,
\end{equation}
and then it's solution will be
\begin{equation} \label{eqNh}
    N^\mu h_{\mu \nu} = h_{2 \nu} \delta(p^2) \, .
\end{equation}
Here, the idea is to see if the remaining gauge freedom is sufficient to fix also the condition $N^\mu h_{\mu \nu} = 0$, or equivalently $h_{2 \nu}= 0$. To see that, we make a new gauge transformation, satisfying \eqref{eqxi} and \eqref{eqcp}, to get in a gauge where
\begin{equation}
    N^\mu h'_{\mu \nu} = 0  = \delta(p^2) (h_{2\nu} +c_\nu+ N \cdot c p_\nu) \, ,
\end{equation}
which has at least the solution
\begin{equation}
    h_{2\nu} +c_\nu+ N \cdot c p_\nu=0 \rightarrow c_\nu=- h_{2\nu} - N \cdot c p_\nu\, ,
\end{equation}
and by observing that $N^\nu c_\nu = - \frac{1}{2} N^\nu h_{2 \nu} $ we conclude
\begin{equation} \label{eqcfin}
    c_\nu=- h_{2\nu} +\frac{1}{2} N^\mu h_{2 \mu} p_\nu\, ,
\end{equation}
condition for which we get to a gauge where we also have $N_\mu h^{\mu \nu}=0$ or $n^\mu h_{\mu \nu}=0$, while preserving all the previous gauge conditions: the Lorentz gauge is maintained because we used for the $\xi_\nu$ of this new gauge transformation the form of \eqref{eqxi}, while the traceless condition is still preserved since \eqref{eqcfin} satisfies \eqref{eqcp} for $h_0 =0$, in fact
\begin{equation}
    c \cdot p =  p^\nu h_{2\nu} +\frac{1}{2} p^2 N^\mu h_{2 \mu} = 0 \,,
\end{equation}
where $p^\nu h_{2\nu} = 0$ due to the Lorentz gauge and $p^2 N^\mu h_{2 \mu}=0$ due to \eqref{eqNh}, demonstrating finally the compatibility of all these gauge conditions.\\

\subsection{Gauge Fixed E.o.M.}
With all that said, we conclude that our E.o.M with gauge conditions 
\begin{eqnarray}
   &&p^\mu h_{\mu \nu}= 0 \, \label{gauge1},\\
   &&n^\mu h_{\mu \nu} = 0 \, \label{gauge2}, \\
   &&h= 0 \label{gauge3}\, ,
\end{eqnarray}
simply becomes a Klein-Gordon equation for the field $h$, that in momentum space reads as the dispersion relation
\begin{equation} \label{KGeq}
(p^2 -m^2_g)h_{\mu \nu} = 0 \, ,
\end{equation}
where $m_g$ plays the role of a mass for the graviton, as anticipated earlier.\\

\section{Solution of the E.o.M.}
At this point, we are ready to find the explicit solution of \eqref{KGeq} in function of the physical degrees of freedom of the theory, which we will demonstrate to be only two. That is in accordance with the fact that, despite the presence of mass, we are not violating gauge invariance.

From \eqref{gauge1} and \eqref{gauge2} we observe that
\begin{equation} \label{h0hi}
    h_{0 \beta} = -\frac{n^i}{n^0} h_{i\beta } = -\frac{p^i}{p^0} h_{i \beta} \rightarrow (\frac{n^i}{n^0}-\frac{p^i}{p^0}) h_{i \beta} = 0 \, ,
\end{equation}
meaning $h_{i\beta}$ will have no projection on the direction $\frac{\vec n}{n^0}-\frac{\vec p}{p^0}$. Therefore, is convenient to choose as a three-dimensional spatial basis the orthogonal set $ \{ \vec u \,,\, \vec M \,,\,  \frac{\vec n}{n^0}-\frac{\vec p}{p^0} \}$, where we have defined the adimensional and $SIM(2)$-invariant vectors
\begin{eqnarray}
   \vec u =  \frac{\vec n}{n^0} \times \frac{\vec p}{p^0}  \; , \;\;\; \vec M =
   \vec u \times ( \frac{\vec n}{n^0}-\frac{\vec p}{p^0}) \, .
   %= 
   %(\frac{\vec p^2}{p_0^2} - \frac{\vec p}{p^0} \cdot \frac{\vec n}{n^0} ) \frac{\vec n}{n^0} +(  \frac{\vec n^2}{n_0^2}-  \frac{\vec n}{n^0} \cdot \frac{\vec p}{p^0}) \frac{\vec p}{p^0} \nonumber 
\end{eqnarray}
Therefore, we can write
\begin{eqnarray} \label{exprh1}
   && h_{i \beta} = A_{\beta} u_i + B_\beta M_i \, , \, \, h_{0 \beta} = -\frac{n^i}{n^0} h_{i \beta} = \frac{n_i}{n_0} h_{i \beta} \, , \nonumber \\
   && h_{ij } =A_j u_i + B_j M_i = A_i u_j + B_i M_j = h_{ji} \, .
\end{eqnarray}
We observe that since we want $( \frac{ n ^i}{n^0}-\frac{ p^i}{p^0}) h_{i j}=0$, then $A_i$, $B_i$ in our basis have to be linear combinations only of $u_i$ and $M_i$
\begin{eqnarray*}
    A_i = a u_i + b M_i \; ,\;\;\; B_i = c u_i + d M_i \, ,
\end{eqnarray*}
where imposing $h_{ij} = h_{ji}$ we find $b=c$. The coefficient $A_0$, $B_0$ are constrained by $h_{0i} = h_{i0}$, from which
\begin{eqnarray*}
    A_0 = \frac{\vec M \cdot \vec n}{n_0} b \;,\;\;\; B_0 =\frac{\vec M \cdot \vec n}{n_0} b \, .
\end{eqnarray*}
Furthermore, by imposing the traceless condition \eqref{gauge3}, we find the expression of $a$ in function of $d$
\begin{equation}
    a = \frac{(\vec M \cdot \vec n)^2 - n_0^2\vec M ^2}{ n_0^2\vec u ^2} d =\tilde a \, d \,.
\end{equation}
Finally, we get the expressions for the components of $h_{\mu \nu}$ in function of the two physical degrees of freedom $b$, $d$
\begin{eqnarray}
    && h_{00}=\left (\frac{\vec M \cdot \vec n}{n_0 } \right )^2 d \;,\;\;\; h_{0i}= \left (\frac{\vec M \cdot \vec n}{n_0} \right ) (b u_i + d M_i) \, , \nonumber \\
    && h_{ij}= \left ( \tilde a u_i u_j +  M_i M_j \right ) d +  (u_i M_j + u_j  M_i) b \, .
\end{eqnarray}

\section{Zero Mass Limit}
We have to verify that, in the limit of zero mass, we recover the physics of linearized GR. To do that, we exploit the gauge invariance property of the linearized Riemann Tensor
\begin{equation}
    R_{\rho \mu \nu \kappa} = \frac{1}{2} ( h_{\rho \nu , \mu \kappa} -h_{\mu \nu , \rho \kappa} -h_{\rho \kappa, \mu \nu} + h_{\mu \kappa , \rho \nu} ) \,.
\end{equation}
Therefore, being gauge invariant, we can prove that, in the massless regime, relations for its components valid in the usual additional gauge $h_{i0}=0$ are still there for our gauge choice, ensuring that the zero mass limit is effectively the linearized known theory of massless gravitons.

\subsection{Temporal and Spatial Components of $ R_{\rho \mu \nu \kappa}$}
Due to its symmetries, in the end, we have just five distinct combinations of spatial or temporal indices we can recognize for the Riemann tensor
\begin{align}
   & R_{0000} = R_{000i} = 0 \, ,\\
   & R_{i00j}=\frac{1}{2} ( h_{0j,0i} +h_{0i,0j}-h_{00,ij} -h_{ij,00}) \,, \label{Rmm}\\
   & R_{0kij}=\frac{1}{2} ( h_{0i,jk} +h_{kj,0i}-h_{0j,ki} -h_{ki,0j}) \, ,\\
   & R_{lkij}=\frac{1}{2} ( h_{li,jk} +h_{kj,li}-h_{lj,ki} -h_{ki,lj}) \, .
\end{align}
In the following, we will use the components of the type \eqref{Rmm} for the application of geodesic deviation. 

\subsection{Reference Choice and Plane Wave Ansatz}
In the classic linearized gravity, for a plane wave propagating in the z-direction, we have that the only non-zero components of $R_{\mu 0 0\nu}$ are $R_{1001}$ and $R_{2002}$, such that $R_{1001}+R_{2002}=0$, and $R_{1002} =R_{2001}$. 

Let's introduce, for our solution, the plane wave ansatz for $h_{\mu \nu}$, labeling the axis of propagation as the z-axis: $p^\mu = (E,0,0,p)$
\begin{equation}
     h_{\mu \nu} =  \mathcal{RE}(A_{\mu \nu} e^{i p^\mu x_\mu})= \mathcal{RE}( A_{\mu \nu} e^{i(E t - p z)}) \, ,
\end{equation}
where $A_{\mu \nu }$ is the polarization tensor satisfying the conditions $p^\mu A_{\mu \nu}= 0 \, , \, \, n^\mu A_{\mu \nu} = 0 \, ,\,\,A^\mu _{\;\mu} = 0$.

Note that, by deriving $h_{\mu \nu}$ respect to $t,z$, we see
\begin{equation} \label{2h0h3}
\begin{cases}
     \partial_0 h_{\mu \nu} = i E h_{\mu \nu} \\ \partial _3 h_{\mu \nu} = -i p h_{\mu \nu}
\end{cases}
    \rightarrow  \; \partial_3 h_{\mu \nu} = - \frac{p}{E} \partial_0 h_{\mu \nu } \, .
\end{equation} 
Furthermore, since here $h_{\mu \nu}$ has no dependence on $x$ and $y$ we have $\partial_1 h_{\mu \nu} = \partial_2 h_{\mu \nu}= 0$, and from the Lorentz gauge condition we find $ p^\mu h_{\mu\nu} = 0 \to  h_{3\nu} = - \frac{E}{p} h_{0\nu}$. \\

\subsection{Riemann Tensor Conditions}
At this point is easy to verify, for example, that in our analytical solution
\begin{align}
    R_{1001}+R_{2002} &= -\frac{1}{2} h_{11,00}-\frac{1}{2} h_{22,00}=  \nonumber\\
    & = -\frac{1}{2}\partial_0^2 (h_{11}+h_{22})= \frac{E^2}{2} (h_{00}-h_{33})= \nonumber\\
    & =\frac{E^2}{2}\frac{(\vec M \cdot \vec n)^2 - n_0^2 M_3^2}{n_0^2} d \propto m^2_g d \, .
\end{align}
Therefore, in the massless limit $R_{1001}+R_{2002} \to 0$. 

The same calculations can be made for the other components of the Riemann tensors, always getting quantities proportional to the VSR mass parameter and then
\begin{eqnarray}
      &&  R_{1001}+R_{2002}\to 0 \,,\\
    && R_{1003} ,  R_{2003} ,  R_{3003} \to 0 \,,
\end{eqnarray}
as expected, due to the gauge invariance of $R_{\rho \mu \nu \kappa}$. 

Note that, defining $\tilde n_i = n_i/n_0 $, the spatial base vectors in the massless limit become
\begin{equation}
    \vec u = (\tilde n^2 , - \tilde n^1, 0) \;,\;\;\; \vec M = (1-\tilde n^3)( \tilde n^1, \tilde n^2, \tilde n^3 + 1) 
\end{equation}

\section{Gravitational Waves and Geodesic Deviation}
As a first application, we want to study the modifications produced by VSR to the known geodesic deviation equations for a gravitational wave, represented by the space-time perturbation $h_{\mu\nu}$.

The expression of the geodesic deviation equation depends on the linearized Riemann Tensor $R_{\mu \nu \alpha \beta}$ in the following way
\begin{equation} \label{geoeq}
    \partial_0^2 \delta \xi^\mu = R^\mu_{\;00\gamma } \delta \xi^\gamma = \eta^{\mu \delta} R_{\delta 00\gamma } \delta \xi^\gamma = \eta^{\mu \mu} R_{\mu 00\gamma } \delta \xi^\gamma \, ,
\end{equation}
The case $\mu = 0$ is trivial, since we have already seen that $R_{000\gamma} = 0$, then $\partial_0 ^2 \xi^0 = 0$, that combined with the initial conditions $\delta \xi^0 (t=0) = \partial_0 \delta \xi^0 (t=0) = 0$, implies $\delta \xi^0 = 0$. So we have no temporal displacement.

For the spatial component of the equation, we see that
\begin{align} \label{geoeqi}
     \partial_0^2 \delta \xi^i &= \eta^{ii} R_{i 00 j } \delta \xi^j = 
     - R_{i 00 j } \delta \xi^j = \nonumber\\
     &= \frac{1}{2} (h_{00,ij } +h_{ij,00} -h_{0i,0j}-h_{0j,0i}) \delta \xi^j \,.
\end{align}
The equation \eqref{geoeqi} for $i=1,2$ becomes
\begin{equation}
    \partial_0^2 \delta \xi^i=  \frac{1}{2} \partial_0^2 h_{i j }\delta \xi^j -\frac{1}{2} \partial_0 \partial_3 h_{0i} \delta \xi^3 \, ,
\end{equation}
but, using \eqref{h0hi} and \eqref{2h0h3}, we get
\begin{equation*}
\partial_0^2 h_{i 3} - \partial_0 \partial_3 h_{0i} = \partial_0^2 h_{i 3} (1 - \frac{p^2}{E^2}) = \frac{m_g^2}{E^2} \partial_0^2 h_{i3} \, .
\end{equation*}
Then
\begin{equation} \label{x1.1}
     \partial_0^2 \delta \xi^i=   \frac{1}{2} \partial_0^2 h_{i 1 }\delta \xi^1 + \frac{1}{2} \partial_0^2 h_{i 2 }\delta \xi^2 + \frac{1}{2} \frac{m_g^2}{E^2} \partial_0^2 h_{i3} \delta \xi^3 \, .
\end{equation}
Since $h_{\mu \nu}$ is a perturbation we can solve the differential equation in a perturbative way by defining $\delta \xi^\mu (t)  = \delta \xi_0^\mu + \delta \xi_1 ^\mu (t)$ where $\delta \xi_1^\mu$ is a small perturbation of $\delta \xi_0^\mu$. 

This way the equation \eqref{x1.1} become:
\begin{equation*} 
     \partial_0^2 \delta \xi^i_1=  \frac{1}{2} \partial_0^2 h_{i 1 }\delta \xi^1_0 + \frac{1}{2} \partial_0^2 h_{i 2 }\delta \xi^2_0 + \frac{1}{2} \frac{m_g^2}{E^2} \partial_0^2 h_{i3} \delta \xi^3_0 \, ,
\end{equation*}
the solution of which, with initial conditions $\delta \xi^i_1 (t= 0) = \partial_0 \delta \xi^i_1 (t= 0) = 0$, is 
\begin{equation*}
     \delta \xi_1^i=  \frac{1}{2}  h_{i 1 }\delta \xi_0^1 + \frac{1}{2}  h_{i 2 }\delta \xi_0^2 + \frac{1}{2} \frac{m_g^2}{E^2} h_{i3} \delta \xi_0^3 \, . 
\end{equation*}
Then, the complete displacement along the $i$ direction is
\begin{equation} \label{eqxii}
    \delta \xi^i= \delta \xi_0^i + \frac{1}{2}  h_{i 1 }\delta \xi_0^1 + \frac{1}{2}  h_{i 2 }\delta \xi_0^2 + \frac{1}{2} \frac{m_g^2}{E^2} h_{i3} \delta \xi_0^3 \, .
\end{equation}
To work out the case $\mu=3$, we follow the same procedure as $\mu =1,2$ but starting from the equation 
\begin{equation*}
     \partial_0^2 \delta \xi^3  = \frac{1}{2} \frac{m_g^2}{E^2} \partial_0 ^2 h_{13} \delta \xi^1 +\frac{1}{2} \frac{m_g^2}{E^2} \partial_0 ^2 h_{23} \delta \xi^2  +\frac{1}{2} \frac{m_g^4}{E^4} \partial_0 ^2 h_{33} \delta \xi^3 \, ,
\end{equation*}
from which, with a little more calculations than before, we get to the final expression for the complete displacement in the $z$-direction
\begin{equation} \label{eqxi3}
     \delta \xi^3 = \delta \xi^3_0 +\frac{1}{2} \frac{m_g^2}{E^2}  h_{13} \delta \xi^1_0 +\frac{1}{2} \frac{m_g^2}{E^2}  h_{23} \delta \xi^2_0  +\frac{1}{2} \frac{m_g^4}{E^4} h_{33} \delta \xi^3_0 \, .
\end{equation}

\subsection{Massless Limit}
First of all, we verify that, in the massless limit, we recover the usual form for the geodesic deviation of the gravitational waves. To do that, is sufficient to demonstrate that $h_{11}+h_{22} \to 0$ for $m_g \to 0$, where
\begin{align}
    h_{11} = (\tilde a u_1^2 + (1+\tilde n_3)^2 \tilde n_1^2) d + 2 \tilde n_1 \tilde n_2 (1+\tilde n_3) b \,,\\
    h_{22} = (\tilde a u_2^2 + (1+\tilde n_3)^2 \tilde n_2^2) d - 2 \tilde n_1 \tilde n_2 (1+\tilde n_3) b\,.
\end{align}
Summing them we obtain
\begin{equation} \label{h11h22}
    h_{11}+h_{22} = (\tilde a + (1+\tilde n_3)^2) (\tilde n_1^2 + \tilde n_2^2) d \,,
\end{equation}
and since $\tilde a = \frac{(\vec M \cdot \tilde n )^2- \vec M^2}{u^2} = - (1+\tilde n_3)^2$, therefore \eqref{h11h22} becomes $ h_{11}+h_{22}= 0$. Then, it's easy to see that, in the massless limit, we reobtain the results of linear GR. \\

\subsection{Massive Effects} 
Letting the mass parameter $m_g$ be different from zero, we see from \eqref{eqxii} and \eqref{eqxi3} three new VSR consequences
\begin{enumerate}
    \item Remarkably, the presence of graviton mass produces a motion also on the propagation direction, in contrast to what happens in the massless case.
    \item Motions along the two transverse directions are modified by the presence of graviton mass.
    \item Hidden in $h_{ij}$ there are anisotropic effects depending on the direction of $\vec n$
\end{enumerate}
The form of these effects depends on the initial condition vector ${\delta \xi_0^\mu}$, but still, their existence is pretty general.

\subsection{VSR Effects' Magnitude}
As we have seen, the corrections to the spatial displacements are proportional to the factor $\frac{m_g^2}{E^2}$. We can give an approximate estimation of this perturbative factor: there exist many different experiments from which we can provide upper bounds to the graviton mass \cite{de2017graviton,shao2020new,will2018solar}. For example, from the time lag measured between the gravitational and electromagnetic signal of the first observed merging of two neutrons stars (GW170817), we can infer a weak upper bound of $m_g \sim 10^{-19} \, eV$ \cite{will2018solar}. Considering other GW detections the upperbound is decreased to $10^{-22}$. From Binary Pulsar \cite{shao2020new} we get an upper bound of $m_g \sim 10^{-28} \, eV$, while from Solar system's tests we get $m_g \sim 10^{-24} \, eV$ \cite{will2018solar}. However, one should not forget that the majority of the upper bound's estimates found in literature for the graviton mass are model dependent.

Therefore, using $10^{-24} \, eV$ as an average upper bound of the graviton mass, in the range of frequencies spanned by the interferometers LIGO and VIRGO, $10 Hz$ to $10 kHz$, the upper bound for our perturbative parameter will be approximately of $\frac{m^2_g}{E^2} \sim 10^{-20}$, making VSR effects probably too small to be detected for our current generation of gravitational wave detectors. 

However, for the future generations of interferometers like LISA \cite{amaro2017laser}, which will explore a lower frequency range $[0.1 mHz, 1 Hz]$, the parameter upper bound increases significantly to $\frac{m^2_g}{E^2} \sim 10^{-10}$, that combined with larger dimensions of future interferometers and the anisotropic nature of VSR could lead to observable effects.

\subsection{Multipolar Nature of Gravitational Radiation}

Due to the gauge invariance of our formulation, we can only couple the perturbation field $h_{\mu \nu}$ to a conserved quantity, which is here represented by the energy-momentum tensor $T_{\mu\nu}$, as usual in gravity theories. The E.o.M. for $ h_{\mu\nu} $, in presence of a source, would therefore relate the quantities
\begin{equation}
    O_{\mu \nu \alpha \beta} h^{\alpha \beta} \propto \, T_{\mu \nu} \, .
\end{equation}
Then, gauge invariance implies
\begin{equation}
   p^\mu O_{\mu \nu \alpha \beta} h^{\alpha \beta} = 0 \to  p^\mu T_{\mu \nu} = 0 \, ,
\end{equation}
which represents the usual equation leading to conservation of total energy $ E $ and momentum $ \vec P $ of the source.

We know that, in GR, the absence of monopolar and dipolar gravitational emission is due to conservation of energy and momentum: let's introduce the gravitational monopole $ M\sim \int d^3x \rho(\vec x) $ and dipole $ \vec D \sim \int d^3x \rho(\vec x) \vec x$, with $\rho(\vec x) $ being the energy density of the system. Thus, since $ M $ and the time derivative $ d_t \vec D$ are proportional to conserved quantities, respectively $ E $ and $ \vec P $, the second time derivative of both the gravitational monopole and dipole is zero $ d^2_t M = d^2_t \vec D= 0 $, leading to a null monopolar and dipolar radiation.

Using the same arguments in VSR we get to the same conclusions as in GR, implying the first non-zero multipolar radiation component for gravitational waves is still the quadrupolar one.

\section{Conclusions}

In this work, we studied the theory of linearized gravity field in the framework of VSR, finding it allows for a graviton mass while preserving gauge invariance. 

We then explored the solution to the new E.o.M. in a specific gauge, with particular attention to the massless limit for which we recover linear GR as expected.

Additionally, we have coupled the linear VSR gravity to matter, in a gauge invariant way,
and verified that dipole radiation of gravitational waves is absent.

Further work must be done to give precise numerical predictions on distinct gravitational phenomena. Nevertheless, the presence of a graviton mass is such an important feature that would affect almost every gravitational area of study, making the VSR massive gravity formulation worth exploring.
%\textbf{Conclusions }--- In this paper, we studied the theory of linearized gravity field in the framework of VSR. Building the model under generic field theory assumptions, we have constructed a coherent approach, which reduces to the linearized General relativity in the limit of $m_g \to 0$. 

%Studying its fundamental consequences for gravitational waves, we ended up predicting two new VSR effects, the magnitude of which could be large enough to make them detectable by the next generation of gravitational interferometers, like LISA, testing in this way the VSR hypothesis.\\

\acknowledgments
A.S. acknowledges financial support from ANID Fellowship CONICYT-PFCHA/DoctoradoNacional/2020-21201387. 
J.A. acknowledges the partial support of the Institute of Physics PUC and Fondo Gemini Astro20-0038.


\nocite{*}

\bibliography{apssamp}% Produces the bibliography via BibTeX.

\end{document}